\documentclass[]{pasj01}
\begin{document}
\Received{2020/4/4}
\Accepted{2020/7/11}
\Published{yyyy/mm/dd}

%\linenumbers

\title{Temporal and Spatial Variation of Synchrotron X-ray Stripes in Tycho's Supernova Remnant}

%%% begin:list of authors
% Do NOT capitalize all letters in "textsc".
\author{Masamune \textsc{Matsuda}\altaffilmark{1}}%
%\thanks{Example: Present Address is xxxxxxxxxx}}
\altaffiltext{1}{Department of Physics, Kyoto University, Kitashirakawa Oiwake-cho, Sakyo, Kyoto, Kyoto 606-8502, Japan}
\email{matsuda.masamune.38a@kyoto-u.jp}

\author{Takaaki \textsc{Tanaka}\altaffilmark{1}}

\author{Hiroyuki \textsc{Uchida}\altaffilmark{1}}

\author{Yuki \textsc{Amano}\altaffilmark{1}}

\author{Takeshi Go \textsc{Tsuru}\altaffilmark{1}}
%%% end:list of authors

%% `\KeyWords{}' always has to be placed before `\maketitle'.
 %Do NOT move this preamble from here!

\maketitle

\begin{abstract}
The synchrotron X-ray ``stripes'' discovered in Tycho's supernova remnant (SNR) have been attracting attention since they may be evidence for proton acceleration up to PeV. 
We analyzed Chandra data taken in 2003, 2007, 2009, and 2015 for imaging and spectroscopy of the stripes in the southwestern region of the SNR.
Comparing images obtained at different epochs, we find that time variability of synchrotron X-rays is not 
limited to two structures previously reported but is more common in the region. 
Spectral analysis of nine bright stripes reveals not only their time variabilities but also a strong anti-correlation between the surface brightness and photon indices. 
The spectra of the nine stripes have photon indices of $\Gamma = 2.1$--$2.6$ and are significantly harder than those of the outer rim of the SNR in the same region with 
$\Gamma = 2.7$--$2.9$. 
Based on these findings, we indicate that the magnetic field is substantially amplified, and suggest that particle acceleration through a stochastic process may be at work 
in the stripes. 
\end{abstract}

%--------------------------------------------------------------------------------

\section{Introduction}\label{sec:intro}
Supernova remnants (SNRs) have long been hypothesized to be major production sites of Galactic cosmic rays up to the so-called 
knee in the cosmic-ray spectrum at $\sim 3~{\rm PeV}$. 
The hypothesis is favored partly because diffusive shock acceleration, expected be working at the expanding shock of SNRs, can naturally 
explain the power-law spectrum of cosmic rays (e.g., \cite{Bla87,Mal01}). 
Energetics is another major reason why SNRs have been regarded as cosmic-ray origin. 
If $\sim 10\%$ of kinetic energy released in supernova explosions is supplied to particle acceleration, SNRs can support the energy density of 
cosmic rays in the interstellar space. 

Detecting synchrotron radiation, X-ray observations have been providing evidence that electrons are accelerated to ultra-relativistic energies 
at blast waves of young SNRs (e.g., \cite{Koy95,Tan08}). 
Gamma-ray emissions, although their radiation mechanisms sometimes are subject to considerable debate, serve as direct evidence 
for particle acceleration in SNRs (e.g., \cite{Aha07}). 
Recent sensitive GeV gamma-ray observations with the Fermi Large Area Telescope and AGILE enabled identification of 
$\pi^0$-decay emission in SNRs, offering a compelling way of probing acceleration of protons and nuclei, the primary component of 
Galactic cosmic rays (\cite{Ack13}; \cite{Jog16}; \cite{Giu11}).  
In spite of the mounting evidence, however, clear indication of acceleration up to the knee has not been obtained from SNRs so far (e.g., \cite{Arc17}). 

Protons should be accelerated up to the knee if SNRs  are indeed the origins of Galactic cosmic rays. 
\citet{Lag83} estimated the maximum energy of particles in SNR shocks as $E_{\rm max} = 100~Z(B/1~\mu{\rm G})~{\rm TeV}$, where $Z$ and $B$ 
are the particle charge and magnetic field strength, respectively. 
The equation implies that the magnetic field should be substantially amplified in order for protons gain energies up to the knee. 
Theoretical studies indeed indicate that cosmic-ray streaming instability can lead to significant amplification of magnetic field at a strong shock (e.g., \cite{Bel01,Bel04}). 
Observationally, thin filamentary structures (e.g., \cite{Bam03,War05}) as well as rapid variability of synchrotron radiation (\cite{Uch07,Uch08,Bor18,Oku20}) 
discovered in X-ray data of SNRs are often interpreted as the results of magnetic field amplification. 

The synchrotron X-ray ``stripes'' in Tycho's SNR discovered by \citet{Eri11} may be indicative of presence of PeV particles as well as magnetic field amplification. 
\citet{Eri11} interpreted that the gaps between the stripes ($\timeform{8''}$) corresponds to twice the gyroradius of accelerated protons, 
concluding that protons reach PeV energies. 
\citet{Oku20} discovered time variable features in the southwestern region of the SNR where stripes are observed. 
The time variability can be explained if the magnetic field is amplified to $\sim 100~\mu{\rm G}$ and/or if magnetic turbulence significantly changes with time. 
The discovery of the stripes triggered some theoretical works proposing models for the peculiar structure (e.g., \cite{Byk11, Mal12, Cap13, Lam15}). 
Yet, their physical origin still is an open question. 

Here we report imaging and spectral studies of the southwestern stripes of Tycho's SNR using Chandra archival data. 
We aim to reveal temporal and spatial variation of synchrotron X-rays of the stripes to discuss their physical origin. 
While \citet{Oku20} focused on two specific features, we characterize spectral and temporal behavior of each stripe in a more systematical way.  
Throughout the paper, the statistical errors are quoted at the $1\sigma$ level.

%--------------------------------------------------------------------------------

\section{Observations and Data Reduction}\label{sec:obs}
We analyzed Chandra ACIS data of Tycho's SNR obtained in 2003, 2007, 2009, and 2015. 
Table \ref{tab:tycho_obs} presents the observation log.
We reprocessed and screened all the data with {\tt chandra\_repro} in CIAO version 4.11 with CALDB version 4.8.2.
The effective exposures after the screening are shown in table \ref{tab:tycho_obs}.
We performed relative astrometric corrections to each dataset. 
We first detected point sources in the field with {\tt wavdetect} in CIAO. 
We then reprojected each events file by cross matching the detected sources with {\tt wcs\_match} and {\tt wcs\_update} tasks of CIAO. 
Low photon statistics did not allow us to detect enough sources to perform the astrometric corrections for observations ${\rm ObsID} = 8551$, 0903, 10904, and 10906. 
We therefore discarded these datasets for the imaging analysis, where accurate corrections are required. 
For the spectral analysis, we used these datasets for better statistics.  
We co-added datasets obtained in the same year. 
The total effective times for the imaging analysis in 2003, 2007, 2009, and 2015  are 146~ks, 109~ks, 634~ks, and 147~ks, respectively, while 
those for the spectral analysis are 146~ks, 142~ks, 734~ks, and 147~ks. 
\begin{table}
	\tbl{Observation log.}{%
	\begin{tabular}{ccc}
		\hline
		ObsID & Start date & Effective exposure \\
		&   & (ks) \\
		\hline
		3837  & 2003-04-29 & 146 \\
		7639  & 2007-04-23 & 109 \\
		8551  & 2007-04-26 & 33 \\
		10093 & 2009-04-13 & 118 \\
		10094 & 2009-04-18 & 90 \\
		10095 & 2009-04-23 & 173 \\
		10096 & 2009-04-27 & 106 \\
		10097 & 2009-04-11 & 107 \\
		10902 & 2009-04-15 & 40 \\
		10903 & 2009-04-17 & 24 \\
		10904 & 2009-04-13 & 35 \\
		10906 & 2009-05-03 & 41 \\
		15998 & 2015-04-22 & 147 \\
		\hline
	\end{tabular}}\label{tab:tycho_obs}
\end{table}
%%%%%%-----------------------------------------------

%%%%%%-----------------------------------------------
\begin{figure*}[t]
	\begin{center}
	\includegraphics[width=160mm]{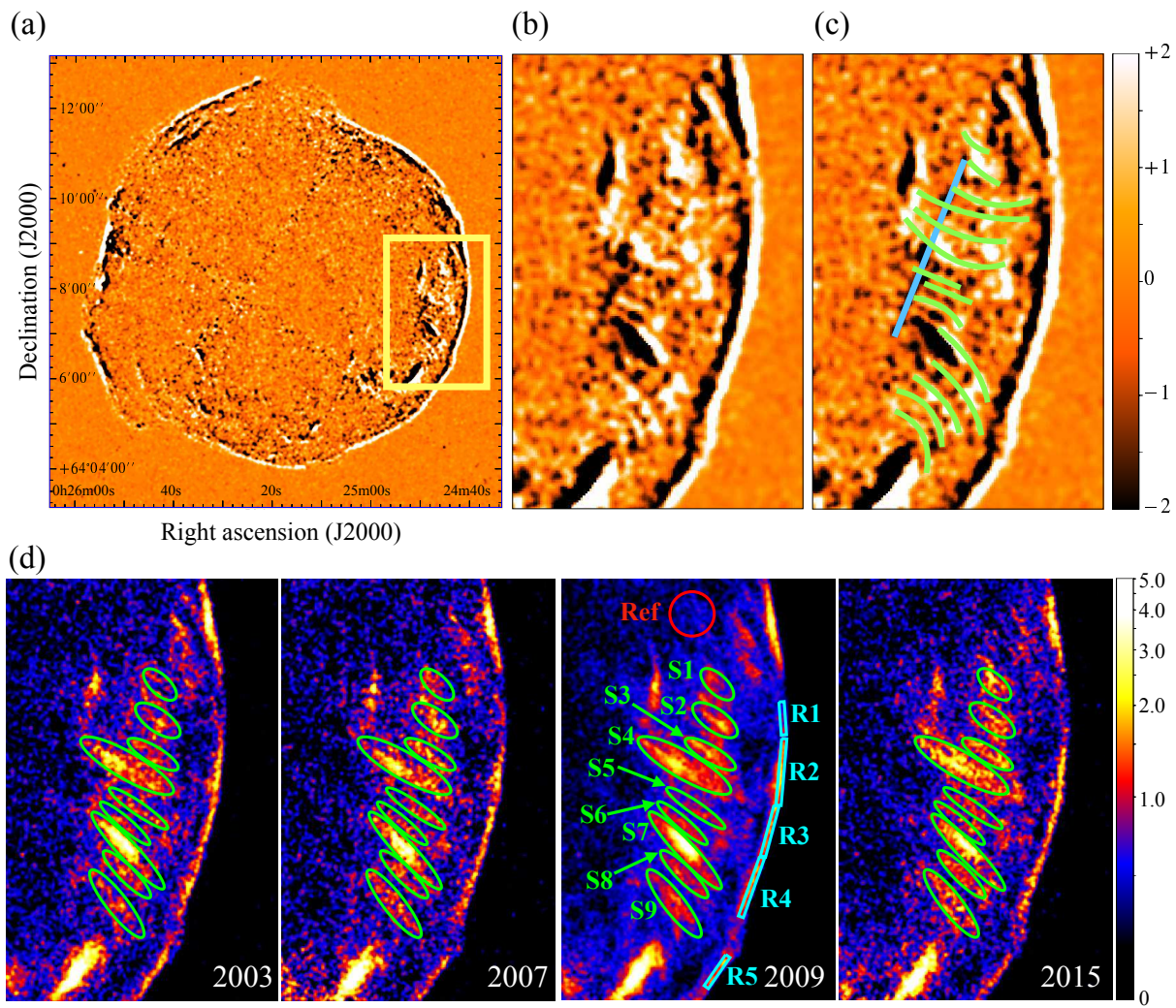}
	\end{center}
	\caption{%
	(a) Difference image between 2003 and 2015 in the 4--6~keV band. 
	The green box corresponds to the region shown in panels (b), (c), and (d).　
	(b) Zoom-in view of the yellow box in panel (a). 
	(c) Same as panel (b) but with guides for identification of notable features overlaid.  
	(d) Exposure-corrected Chandra ACIS images of Tycho's SNR in 2003, 2007, 2009, and 2015.
	The energy band is 4--6~keV.
	The regions S1--S9 and R1--R5 are those for the spectral analysis.
	The Ref region is used for estimating the parameters for the thermal component of the emissions of the stripes.  
	In all the panels, the unit for the color scale is $10^{-8}~{\rm ph}~{\rm s}^{-1}~{\rm cm} ^{-2}$.}%
	\label{fig:TychoAnaReg}
\end{figure*}
%%%%%%-----------------------------------------------

%--------------------------------------------------------------------------------
\section{Analysis and Results}\label{sec:ana}
\subsection{Imaging Analysis}\label{sec:imgana}
%---------------------------------
\begin{figure}[tb]
	\begin{center}
	\includegraphics[width=80mm]{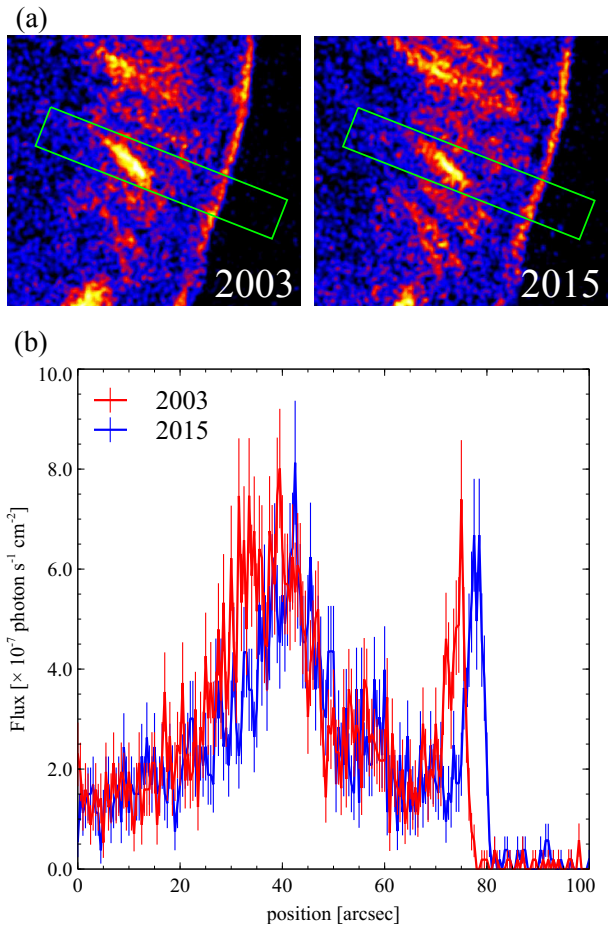}
	\end{center}
	\caption{%
	(a) Regions used for the projection in panel (b). 
	(b) Profiles extracted from the regions shown in panel (a).
	The widths of each bin are \timeform{0.5''}. The geometrical center of the SNR is to the left.}%
	\label{fig:TychoProfReg}
\end{figure}
%---------------------------------
%---------------------------------
\begin{figure}[thb]
	\begin{center}
  	\includegraphics[width=80mm]{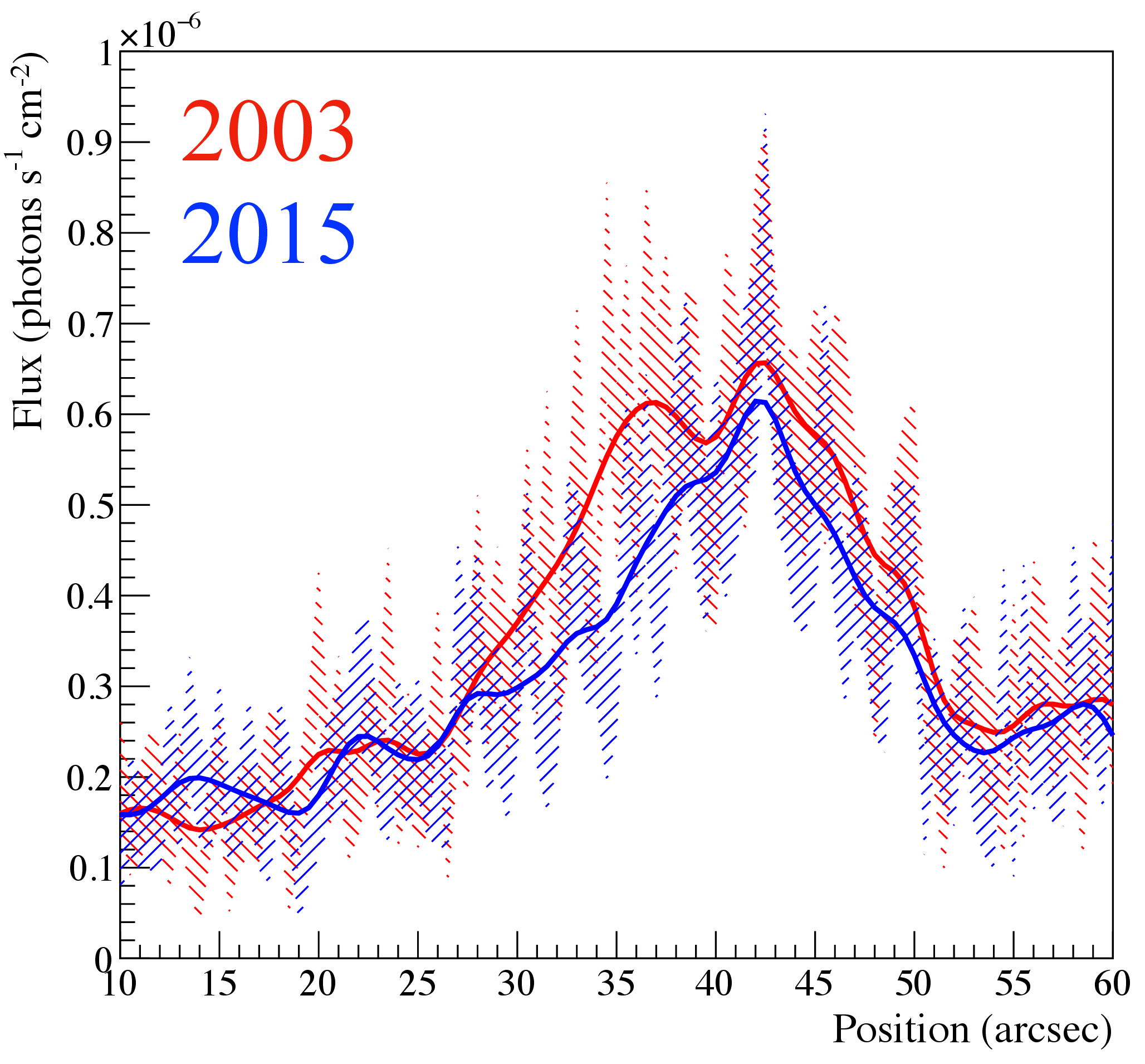}
	\end{center}
	\caption{%
	Comparison between the profile in 2015 and the moved profile in 2003 by \timeform{3.0''}.
	Oblique lines correspond to the range of 1$\sigma$.
	Solid lines show smoothed profiles with bandwidth$=$3.}%
	\label{fig:SmoothedProf}
\end{figure}
%---------------------------------
Figure~\ref{fig:TychoAnaReg}a shows a difference image of Tycho's SNR constructed by subtracting an exposure-corrected image taken 
in 2003 from that in 2015. 
A zoom-in view of the western region is given in figure~\ref{fig:TychoAnaReg}b. 
We also present exposure-corrected images from each year in figure~\ref{fig:TychoAnaReg}d. 
We selected an energy band of 4--6~keV, where synchrotron emission is dominant (e.g., \cite{Eri11}). 
Most of the stripe features appear to move outward as the outer rim, which most probably corresponds to the blast waves \citep{War05}, expands. 
However, some of the changes cannot simply be accounted for just by the expansion. 
The stripes along the green curves in figure~\ref{fig:TychoAnaReg}c generally become brighter in 2015. 
We note that most of the brightening stripes are too faint to be seen in the exposure-corrected images in figure~\ref{fig:TychoAnaReg}d. 
On the other hand, the emission along the cyan line in figure~\ref{fig:TychoAnaReg}c shows a flux decrease. 
These results indicate that the small knot structure, which \citet{Oku20} reported, is not the only structure that
shows brightening over the years in this region. 
We also found hints of the proper motions of the faint stripes along the direction perpendicular to the shock normal. 

Figure~\ref{fig:TychoProfReg}b shows projections along the azimuthal direction from the rectangular region in figure~\ref{fig:TychoProfReg}a, which 
includes the brightest stripe (S7 in figure~\ref{fig:TychoAnaReg}c). 
The sharp and broad peaks correspond to the rim and the stripe, respectively. 
Both peaks clearly are moving outward with time as already suggested by the difference image in figure~\ref{fig:TychoAnaReg}a. 
We first measured the proper motion of the rim. 
Artificially shifting the profile in 2003, we calculated $\chi^2$ between the shifted profile in 2003 and the observed profile in 2015 in the range of \timeform{66''}--\timeform{90''}, 
and searched for a shift that gives the minimum $\chi^2$. 
As a result, we obtained a velocity of $0.29\pm0.01~{\rm arcsec}~{\rm yr}^{-1}$, which can be translated into $3400\pm100~{\rm km}~{\rm s}^{-1}$ 
with a distance of 2.5~kpc assumed \citep{Zhou2016}. 
We note that \citet{Wil16} reported a proper motion consistent with ours (The region we analyzed roughly coincides with their Reg~13).
Measurement of the proper motion of the stripe, on the other hand, was found to be difficult because of its time-variable shape. 
A closer look at the profiles plotted in figure~\ref{fig:TychoProfReg}b suggests that the peak corresponding to the stripe becomes narrower in 2015 than in 2003. 
This is more clearly visible in smoothed profiles presented in figure~\ref{fig:SmoothedProf}. 
In this figure, we shifted the profile in 2003, assuming a velocity of $0.29~{\rm arcsec}~{\rm yr}^{-1}$, which is the measured value for the rim. 
The locations of the peaks are roughly consistent with each other, leading to a conclusion that the proper motion of the stripe is at the same level as the rim.

%--------------------------------------------------------------------------------

\subsection{Spectral Analysis}\label{sec:specana}
%%%%%%-----------------------------------------------
\begin{table*}
\begin{center}
  \tbl{Best-fit parameters in 2009}{%
  \scalebox{0.85}{%
  \begin{tabular}{lccccccccc}
    \hline %-----------------------------------------------
    Parameters~(unit)
     & S1 & S2 & S3 & S4 & S5 & S6 & S7 & S8 & S9\\
    %
    %-----------------------------------------------
    Solid angle~$(\mathrm{arcsec^{2}})$
     & 143.8 & 210.2 & 172.7 & 505.3 & 161.7 & 141.1 & 338.8 & 183.4 & 311.6\\
    \hline %-----------------------------------------------
    $N_\mathrm{H}$~$(\mathrm{10^{21}~cm^{-2}})$
     & $6.80\pm0.17$ & $6.37^{+0.26}_{-0.10}$ & $5.23^{+0.21}_{-0.12}$ & $5.83^{+0.05}_{-0.06}$ & $5.92^{+0.24}_{-0.20}$ & $6.32^{+0.30}_{-0.19}$ & $6.03^{+0.17}_{-0.12}$ & $6.56^{+0.20}_{-0.28}$ & $5.26^{+0.14}_{-0.12}$ \\
     %
    %-----------------------------------------------
    \underline{Power law} \\
    $\Gamma$
     & $2.39\pm0.04$ & $2.37^{+0.04}_{-0.03}$ & $2.06\pm0.03$ & $2.15\pm0.01$ & $2.44\pm0.03$ & $2.40\pm0.04$ & $2.12^{+0.02}_{-0.03}$ & $2.25^{+0.02}_{-0.03}$ & $2.55\pm0.02$\\
    $\mathrm{Flux}$\footnotemark[$*$]
     & $0.43\pm0.01$ & $0.77\pm0.01$ & $0.76\pm0.01$ & $2.35\pm0.02$ & $0.50\pm0.01$ & $0.46\pm0.01$ & $2.08^{+0.02}_{-0.01}$ & $0.64\pm0.01$ & $1.05^{+0.01}_{-0.02}$\\
    \hline %-----------------------------------------------
    \underline{IME compconent} \\
    \begin{tabular}[c]{@{}l@{}}
    $\mathrm{Norm.}$\footnotemark[$\dag$] \\
    $(10^{9}~\mathrm{cm^{-5}})$
    \end{tabular}
     & $3.31^{+0.42}_{-0.38}$ & $1.48^{+0.41}_{-0.43}$ & $1.71^{+0.24}_{-0.36}$ & $4.97^{+0.86}_{-0.26}$ & $2.11^{+0.70}_{-0.40}$ & $1.51^{+0.54}_{-0.38}$ & $1.50^{+0.49}_{-0.61}$ & $2.22^{+0.50}_{-0.42}$ & $0.35^{+0.06}_{-0.04}$\\
    $kT_\mathrm{e}~(\mathrm{keV})$
     & $1.49^{+0.04}_{-0.03}$ & $1.65^{+0.03}_{-0.05}$ & $1.68^{+0.01}_{-0.04}$ & $1.55\pm0.01$ & $1.58^{+0.03}_{-0.02}$ & $1.57^{+0.04}_{-0.03}$ & $1.71^{+0.03}_{-0.04}$ & $1.45^{+0.01}_{-0.04}$ & $1.41\pm0.03$\\ 
    \begin{tabular}[c]{@{}l@{}}
    $n_\mathrm{e}t$ \\
    $(\mathrm{10^{10}~s~cm^{-3}})$
    \end{tabular} & \multicolumn{9}{c}{$4.53~\mathrm{(fixed)}$} \\
    $\mathrm{[Mg/C]/[Mg/C]_{\odot}}$
     & $4.8^{+0.6}_{-0.4}$ & $7.2\pm1.4$ & $9.2^{+1.9}_{-0.8}$ & $8.1^{+0.4}_{-0.7}$ & $5.2^{+0.3}_{-1.3}$ & $6.4^{+1.7}_{-1.6}$ & $10.6^{+6.9}_{-2.1}$ & $5.5^{+0.8}_{-0.9}$ & $4.8\pm0.6$ \\ 
    $\mathrm{[Si/C]/[Si/C]_{\odot}}$
     & $84^{+10}_{-9}$ & $161^{+68}_{-35}$ & $190^{+40}_{-27}$ & $238^{+23}_{-22}$ & $183^{+41}_{-37}$ & $189^{+38}_{-46}$ & $279^{+194}_{-69}$ & $112^{+33}_{-20}$ & $73^{+10}_{-9}$\\
    $\mathrm{[S/C]/[S/C]_{\odot}}$
     & $77^{+10}_{-8}$ & $130^{+67}_{-28}$ & $143^{+31}_{-19}$ & $185^{+11}_{-20}$ & $134^{+30}_{-28}$ & $139^{+39}_{-36}$ & $194^{+144}_{-48}$ & $91^{+25}_{-16}$ & $81\pm11$\\ 
    $\mathrm{[Ar/C]/[Ar/C]_{\odot}}$
     & $85^{+14}_{-11}$ & $105^{+40}_{-12}$ & $110^{+13}_{-10}$ & $175^{+8}_{-11}$ & $99^{+24}_{-9}$ & $106^{+31}_{-22}$ & $202^{+140}_{-65}$ & $77\pm9$ & $86^{+14}_{-13}$\\ 
    $\mathrm{[Ca/C]/[Ca/C]_{\odot}}$
     & $156^{+28}_{-25}$ & $202^{+72}_{-26}$ & $158^{+55}_{-25}$ & $163^{+20}_{-29}$ & $159^{+24}_{-23}$ & $258^{+86}_{-77}$ & $298^{+93}_{-77}$ & $142\pm24$ & $168^{+33}_{-30}$\\
     %
    %-----------------------------------------------
    \underline{Fe component} \\
   \begin{tabular}[c]{@{}l@{}}
    $\mathrm{Norm.}$\footnotemark[$\dag$] \\
    $(10^{9}~\mathrm{cm^{-5}})$
    \end{tabular}
     & $3.31$\footnotemark[$\dag$] & $1.48$\footnotemark[$\dag$] & $1.77$\footnotemark[$\dag$] & $4.97$\footnotemark[$\dag$] & $2.11$\footnotemark[$\dag$] & $1.51$\footnotemark[$\dag$] & $1.50$\footnotemark[$\dag$] & $2.22$\footnotemark[$\dag$] & $0.37$\footnotemark[$\dag$]\\
    $kT_\mathrm{e}$~$(\mathrm{keV})$
     & $5.54^{+0.91}_{-0.83}$ & $4.30^{+1.30}_{-0.98}$ & $3.85^{+1.20}_{-0.80}$ & $4.63^{+0.40}_{-0.26}$ & $2.36^{+0.35}_{-0.39}$ & $3.00^{+1.84}_{-0.83}$ & $2.04^{+0.62}_{-0.26}$ & $2.41^{+0.26}_{-0.55}$ & $1.53^{+0.12}_{-0.23}$\\
    \begin{tabular}[c]{@{}l@{}}
    $n_\mathrm{e}t$ \\
    $(\mathrm{10^{10}~s~cm^{-3}})$
    \end{tabular} & \multicolumn{9}{c}{$0.74~\mathrm{(fixed)}$} \\ 
    \begin{tabular}[c]{@{}l@{}}
    $\mathrm{[Fe/C]/[Fe/C]_{\odot}}$ \\
    $(=\mathrm{[Ni/C]/[Ni/C]_{\odot}})$
    \end{tabular} &
     $3.6^{+0.5}_{-0.4}$ & $5.4^{+1.9}_{-1.2}$ & $5.4^{+1.4}_{-0.5}$ & $9.2^{+0.8}_{-1.0}$ & $7.6^{+1.4}_{-1.1}$ & $5.5\pm1.2$ & $8.0^{+3.2}_{-1.5}$ & $3.8^{+0.8}_{-0.5}$ & $4.8^{+0.6}_{-0.5}$\\
     %
    %-----------------------------------------------
    \underline{Gaussian} \\
    $\mathrm{Norm.}$\footnotemark[$\ddag$]
     & $0.53\pm0.05$ & $0.33\pm0.05$ & $0.36\pm0.04$ & $1.40^{+0.08}_{-0.07}$ & $0.60\pm0.05$ & $0.47\pm0.05$ & $0.44^{+0.06}_{-0.07}$ & $0.35^{+0.04}_{-0.05}$ & $0.61^{+0.08}_{-0.07}$\\
    $\mathrm{Centroid}$~$\mathrm{(keV)}$
     & $1.25^{+0.02}_{-0.01}$ & $1.28^{+0.05}_{-0.02}$ & $1.24\pm0.01$ & $1.253^{+0.006}_{-0.001}$ & $1.238^{+0.002}_{-0.007}$ & $1.234\pm0.003$ & $1.232^{+0.022}_{-0.002}$ & $1.254^{+0.007}_{-0.005}$ & $1.23$ (fixed) \\
    \hline %-----------------------------------------------
    $\chi^{2}~(d.o.f.)$
     & $405~(291)$ & $375~(331)$ & $463~(338)$ & $1153~(433)$ & $593~(302)$ & $451~(299)$ & $598~(418)$ & $392~(321)$ & $606~(373)$\\
    \hline %-----------------------------------------------
    \end{tabular}}%scalebox
    }%tbl
    \label{tab:fitresult}
\end{center}
\begin{tabnote}
\footnotemark[$*$]
The photon flux in the energy band of 4--6~$\mathrm{keV}$.
 The unit is $\times 10^{-13}~\mathrm{erg~s^{-1}~cm^{-2}}$. \\
\footnotemark[$\dag$]
Emmission measures of Fe and IME component are defined by $\int n_{\mathrm{e}}n_{\mathrm{C}}dV/4\pi d^{2}\cdot\mathrm{[C/H]_{\odot}}$ and linked to each other. \\
\footnotemark[$\ddag$]
The unit is $\times \mathrm{10^{-5}~photons~s^{-1}~cm^{-2}}$
\end{tabnote}
\end{table*}

\begin{figure*}[tbp]
   \begin{center}
   \includegraphics[width=160mm]{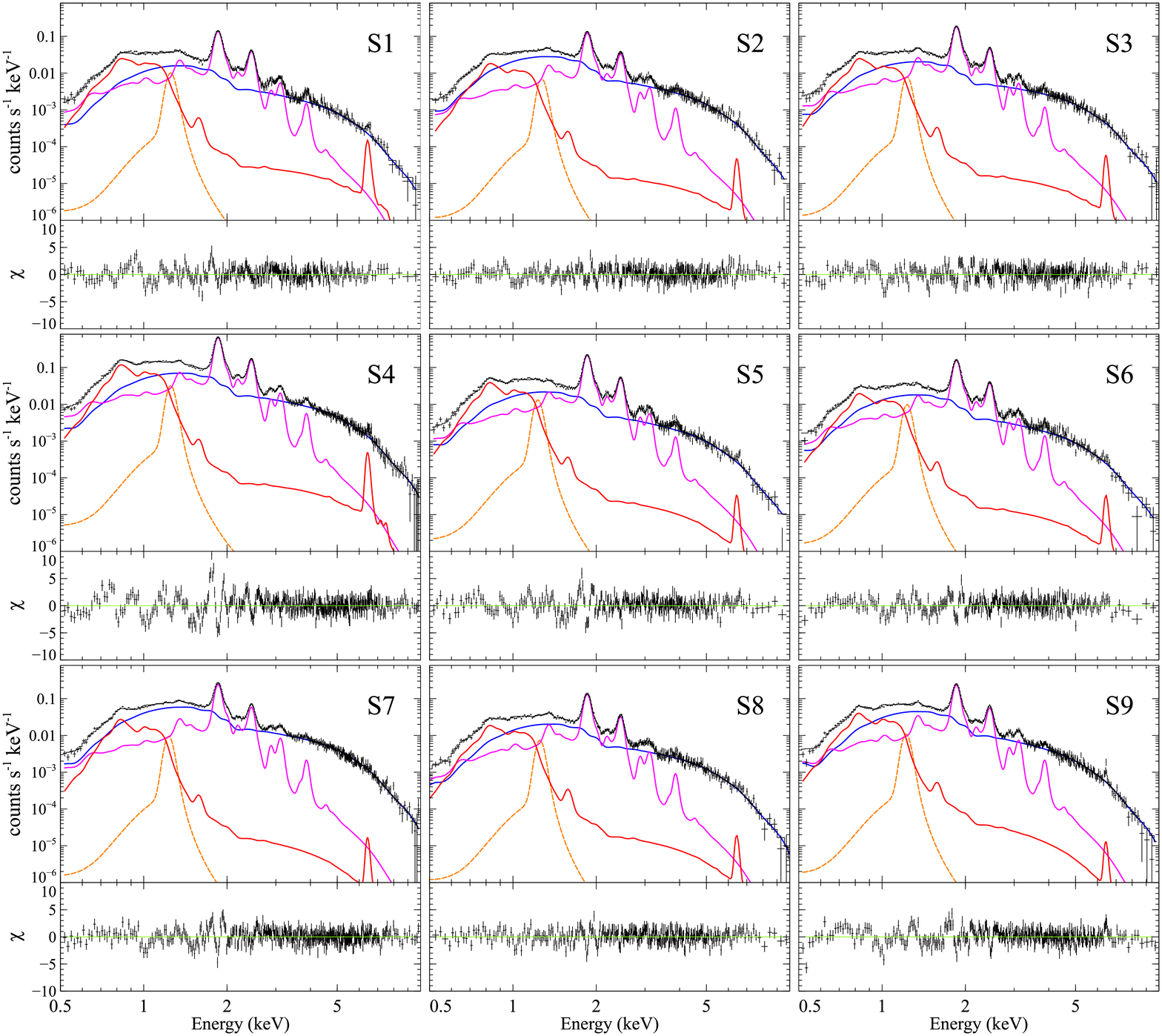}
   \end{center}
   \caption{%
   Spectra and the best-fit models for regions S1--S9 obtained from the data taken in 2009.
   The black curves are the sums of all the components whereas the other colors show contributions from each component. 
   The red and orange curves represent the NEI models for the Fe and IME components, respectively.
   The magenta dashed curves indicate the Gaussian added to the model.
   The model curves for the power-law component are drawn in blue. 
   }%
   \label{fig:Spectrum}
\end{figure*}
%%%%%%-----------------------------------------------
We performed spectroscopy of each bright stripe by extracting spectra from the nine regions labeled as S1--S9 in figure~\ref{fig:TychoAnaReg}c. 
Backgrounds are extracted from the outside of the SNR within the ACIS-I array.
The spectra were binned so that each bin has at least 10 counts and were fitted over the 0.5--10~keV energy band with XSPEC 12.10.1f \citep{Arn96}.
Following the works by \citet{Sat17}, \citet{Yam17}, and \citet{Oku20}, we fitted the spectra with a model consisting of non-thermal and thermal components.
We modeled the non-thermal component with a power law. 
To the thermal emission, which can be attributed to the supernova ejecta, we applied a two-component non-equilibrium ionization (NEI) model using the {\tt vnei} model in XSPEC.
We employed the Tuebingen-Boulder absorption model (\texttt{TBabs}; \cite{Wil00}) for interstellar absorption.

One of the two NEI components represents the emission from Fe whereas the other is for intermediate-mass elements (IMEs: Mg, Si, S, Ar, and Ca).
We treated the abundances of IMEs and Fe as free parameters, and linked the abundance of Ni to Fe.
Since Tycho's SNR is of type Ia origin, the abundances of H, He, and N were fixed to zero.
The abundances of O and Ne with respect to C were fixed at the solar values because C has the lowest atomic number in the elements that would be present in the ejecta.
Emission measures of the Fe and IME components ($\equiv\frac{1}{\mathrm{[C/H]_{\odot}}}\frac{1}{4\pi d^2}\int n_{\mathrm{e}}n_{\mathrm{C}}\, dV$) were linked to each other, 
where $d$ is the distance to Tycho's SNR, $n_{\mathrm{e}}$ and $n_{\mathrm{C}}$ are number densities of electron and carbon, and $V$ is the volume of the emitting plasma.
Fitting the spectra, we found residuals at $\sim 1.2~{\rm keV}$, which are also seen in spectra of Tycho's SNR (e.g., \cite{Sat17, Oku20}) as well as 
other SNRs (e.g., \cite{Oko19}). 
Although the cause of the residuals is not clear (see a discussion by \cite{Oko19}), we added a Gaussian to the model to improve the fits. 
The centroid energy of the Gaussian was allowed to vary for spectra from all the regions except for S9. 
We fixed it to 1.23~keV for S9 since it cannot be well constrained. 

Strong non-thermal emission of the stripes makes it difficult to determine the parameters for the NEI components. 
To constrain the parameters, therefore, we analyzed a spectrum extracted from the region with less contribution from the non-thermal component, which 
is labeled as ``Ref'' in figure~\ref{fig:TychoAnaReg}c. 
Since the region is located at a similar radius of the SNR to stripes, we assumed that ionization ages ($n_e t$) of the NEI components are 
common between the ``Ref'' region and stripes, and determined them by fitting the ``Ref'' spectrum. 
The spectral fitting yielded $n_{\mathrm{e}}t = 4.52\times10^{10}$ s cm$^{-3}$ and $n_{\mathrm{e}}t = 0.74\times10^{10}$ s cm$^{-3}$ for the IME and 
Fe components, respectively. 
We fixed $n_{\mathrm{e}}t$ to these values when fitting the spectra of the stripe regions. 

\begin{figure}[tb]
	\begin{center}
  \includegraphics[width=80mm]{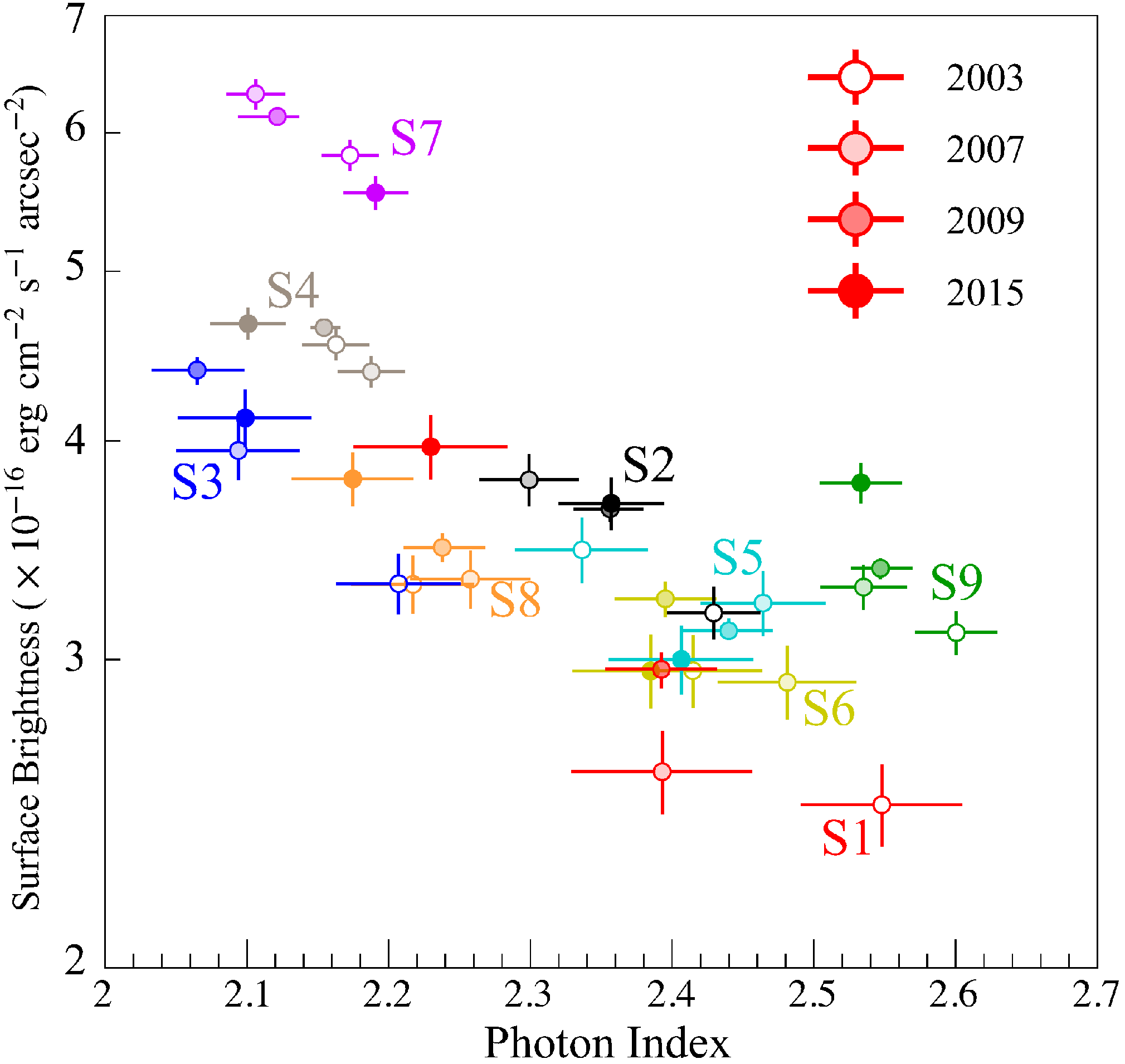} 
  \end{center}
  \caption{%
  Relation between the surface brightness and the photon indices in the stripes.
  The data points from each stripe are plotted in different colors. 
  The color tint indicates the epochs of the data points. 
  }%
  \label{fig:SurBri-PhtnIdx:4yr}
\end{figure}

We first fitted spectra from observations in 2009, which have the highest statistics thanks to the longest exposure time. 
The spectra are plotted with the best-fit models in figure~\ref{fig:Spectrum}, and  
the best-fit parameters are summarized in table~\ref{tab:fitresult}. 
We then fitted spectra from 2003, 2007, and 2015 observations to see time variability of the stripe emissions. 
The parameters for the thermal components were fixed to those obtained for the spectra in 2009 except for the emission measures. 
In figure~\ref{fig:SurBri-PhtnIdx:4yr}, we plot surface brightness of the non-thermal component as a function of photon index, 
which reveals a significant stripe-to-stripe variation of the parameters as well as time variability of each stripe. 
Another finding to note here is the strong anti-correlation between surface brightness and photon indices of the stripe emission.

\section{Discussion}\label{sec:disc}
Our imaging and spectral analyses have revealed time variable nature of the stripes in Tycho's SNR. 
Although \citet{Oku20} reported time variability only of two structures, including S1 in our definition, 
similar variability seems to be rather universal in this part of the SNR. 
Figure~\ref{fig:SurBri-PhtnIdx:4yr} indeed indicates significant flux variability of S2, S3, S7, and S9, in addition to S1. 
The fluxes of the stripes S2 and S3 are lower in 2003 than in the other years. 
The stripe S7 underwent a flux brightening from 2003 to 2007 and then decayed. 
In the case of S9, a continuous flux increase was observed from 2003 to 2015. 
The images in figure~\ref{fig:TychoAnaReg} furthermore indicate not only the bright stripes but also much fainter structures are also variable. 

Following \citet{Uch07}, \citet{Uch08}, and \citet{Oku20}, we can estimate the magnetic field strength of the emitting region 
if we attribute the brightening to production of relativistic electrons through acceleration and the flux decay to decrease of electrons 
emitting synchrotron X-rays. 
Assuming diffusive shock acceleration, we can write the acceleration timescale as 
\begin{eqnarray}
t_{\mathrm{acc}}&=&4\eta\left(\frac{\varepsilon}{\mathrm{keV}}\right)^{0.5}\left(\frac{B}{400~\mathrm{\mu G}}\right)^{-1.5}\left(\frac{v_{\mathrm{sh}}}{3400~\mathrm{km~s^{-1}}}\right)^{-2}~\mathrm{yr}, 
\end{eqnarray}
where $\eta~(\geq 1)$ is the so-called ``gyrofactor'', $\varepsilon$ is synchrotron photon energy, $B$ is the magnetic field strength, and $v_{\mathrm{sh}}$ is the shock velocity. 
We assumed $v_{\mathrm{sh}} = 3400~{\rm km}~{\rm s}^{-2}$  according to our proper motion measurement in \S\ref{sec:imgana}. 
The synchrotron cooling timescale can be given as 
\begin{eqnarray}
t_{\mathrm{syn}}&=&4\left(\frac{\varepsilon}{\mathrm{keV}}\right)^{-0.5}\left(\frac{B}{500~\mathrm{\mu G}}\right)^{-1.5}~\mathrm{yr}. 
\end{eqnarray}
Since we observed flux changes in a timescale of several years, the above equations lead to a conclusion that the magnetic field 
in the stripe region is $\sim 500~\mu {\rm G}$. 
We note here that this estimate would have some uncertainties. 
The above equations assume that electrons with a certain energy emit monochromatic synchrotron photons, 
which is not true in reality and hence produces uncertainties.

The 3D location of the stripes would be key information for discussing their physical origin. 
In the above magnetic field strength estimation, we implicitly assumed that the stripes are on the projected blast waves of the SNR. 
In this case, proper motion of the stripes should appear to be slower than the expansion velocity since only the transverse velocity component is observed. 
The transverse velocity is calculated to be $0.25~{\rm arcsec}~{\rm yr}^{-1}$ , assuming spherical shell expansion with a velocity of $0.29~{\rm arcsec}~{\rm yr}^{-1}$ (\S\ref{sec:imgana}). 
We found the difference of the two values is too small to be measured with the present data, considering the fact that the morphology 
of the stripe is also changing with time (figure~\ref{fig:SmoothedProf}). 
On the other hand, if the stripes are located inside the shell, or far downstream the blast waves, 
the transverse velocity of the stripes can be different from the above case. 
The proper motion of some structures perpendicular to the shock normal (figure~\ref{fig:TychoAnaReg}b) would be in favor of this scenario. 
Models proposed in literature place the stripes in different locations (see e.g., \cite{Byk11}; \cite{Mal12}; \cite{Cap13}; \cite{Lam15}). 
%Models proposed in literature place the stripes in different locations: \citet{Mal12} and \citet{Cap13} assumed that the stripes are in the upstream 
%or in the shock precursor, whereas \citet{Byk11} and \citet{Lam15} claimed that they are in the downstream region. 
A future Chandra observation at another epoch would be needed to measure the proper motion of the stripes precisely enough to give meaningful comparison
with that of the blast waves, which is a key to pinning down the line-of-sight locations of the stripes. 

\begin{figure}[tb]
   \begin{center}
   \includegraphics[width=80mm]{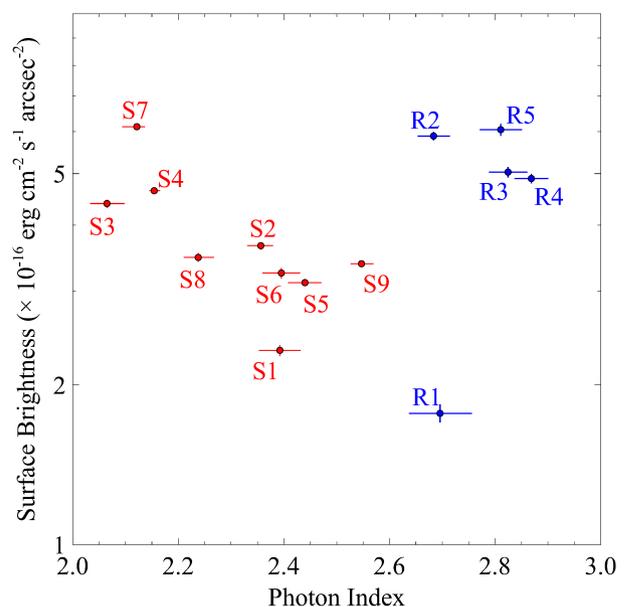} 
   \end{center}
   \caption{%
   Comparison of the relation  between the stripes (red) and the rim (blue).
   The data points plotted here are from the observations in 2009. 
   }%
   \label{fig:SurBri-PhtnIdx:2009}
\end{figure}

Our spectroscopy of each stripe at each epoch has revealed anti-correlation between surface brightness and photon indices (figure~\ref{fig:SurBri-PhtnIdx:4yr}). 
In order to see whether or not the synchrotron-dominant rim emission has a similar anti-correlation as well, we extracted spectra from regions R1--R5 
defined in figure~\ref{fig:TychoAnaReg}c. 
Since thermal emissions are negligible in these regions, we fitted the spectra with an absorbed power law and plotted 
the result in figure~\ref{fig:SurBri-PhtnIdx:2009} together with those from the stripes as observed in 2009. 
In contrast to the stripes, the data points from the rim do not show a significant anti-correlation. 
Also, the rim emission is softer than the stripes with photon indices of $\Gamma = 2.7$--2.9 as compared to $\Gamma = 2.1$--2.6 of the stripes. 
This is consistent with the result by \citet{Lop15}, who found the hardest $> 10~{\rm keV}$ emission with NuSTAR in the west of the SNR coinciding with the location of the stripes.

Then what makes the spectra of stripes harder? 
Since the X-ray band corresponds to the cutoff region of a synchrotron spectrum, 
photon indices reflect the cutoff energy ($\varepsilon_0$): a harder spectrum means a higher $\varepsilon_0$ and vice versa. 
Let us first discuss the case in which stripes are associated with structures of the blast wave region although distinct spectra characteristics 
between the stripes and blast waves (figure~\ref{fig:SurBri-PhtnIdx:2009}) make this case less likely. 
According to the NuSTAR result by \citet{Lop15}, the highest photon energy of synchrotron emission in Tycho's SNR seems to be limited by its age ($=$ acceleration time). 
Given that, the cutoff energy depends both on the shock velocity and on the magnetic field strength as $\varepsilon_0 \propto {v_{\mathrm{sh}}}^4 B^3$ \citep{Lop15}. 
Thus, the hard spectra of the stripes can be ascribed to fast shock velocity and/or strong magnetic field of the region. 
Considering the peculiar morphology of the stripes, it would be rather unlikely that only a fast shock velocity accounts for the hardness. 
Instead, it would be more probable that the magnetic field is amplified in the stripes through, for example, the resonant \citep{Ski75} or 
non-resonant \citep{Bel04} cosmic-ray streaming instability. 
The problem about this scenario is that we cannot explain the short time variability of the stripes at the same time. 
The timescale of the variability expected in the age-limited case would be in the order of the age of the SNR, $\sim 100~\mathrm{yr}$, which is much longer than 
observed. 
One of the possible solutions that can reconcile with both the result by \citet{Lop15} and the variable stripe emissions would be that synchrotron emissions 
of most of the regions are age-limited whereas those right at the stripes are loss-limited with the amplified magnetic field.

If the stripes are not projections of blast waves but are located far downstream of the shock, the hard spectra are somewhat puzzling. 
After being accelerated at the blast wave, electrons are transported downstream through advection or diffusion. 
While being transported, ultra-relativistic electrons lose their energies via severe synchrotron cooling loss, which makes the 
electron spectrum and thus synchrotron X-ray spectrum softer. 
Such softening is indeed observed in Tycho's SNR by \citet{Cas07} with Chandra. 
One of the possible mechanisms to make the synchrotron spectra harder in the stripes is boosting synchrotron photon energy with a strong 
magnetic field. 
Since synchrotron photon energy ($\varepsilon$) is related to parent electron energy ($E_\mathrm{e}$) as $\varepsilon \propto B {E_\mathrm{e}}^2$, 
a stronger magnetic field can make the synchrotron cutoff energy higher and thus synchrotron spectra observed with Chandra harder. 
In addition, if compressible waves/turbulence is present in the stripes, stochastic acceleration may occur and electron spectra would become even harder. 
We note that \citet{Zha15} theoretically studied such a scenario.

In either case of the two discussed above, it would be very challenging to explain also the anti-correlation between surface brightness and photon indices 
(figure~\ref{fig:SurBri-PhtnIdx:4yr}). 
The result would indicate only a small number of parameters are responsible for the temporal and spatial variation of the stripe emission 
otherwise such a tight anti-correlation would not appear. 
Another fact to note about the result in figure~\ref{fig:SurBri-PhtnIdx:4yr} is that the surface brightness is similar between stripes. 
This suggests that line-of-sight depths of the stripes are similar to each other, under an assumption that the nine stripes have similar magnetic field strengths 
and relativistic electron densities. 
If so, it would be more natural to consider that the stripes are shaped like spheroids rather than thin sheets.

%-----------------------------------------------
%-----------------------------------------------

\section{Conclusions}\label{sec:conc} 
Using Chandra data obtained in 2003, 2007, 2009, and 2015, 
we searched for temporal and spatial variation of synchrotron radiation of the stripes in the southwestern region of Tycho's SNR 
in a more systematical way than the work by \citet{Oku20}. 
Our imaging analysis revealed time variability of the emission in this region. 
Analyzing spectra of nine bright stripes, we found significant time variabilities not only of the stripe S1 previously reported by \citet{Oku20} but also of other stripes S2, S3, S7, and S9.
If we attribute the flux increase to production of X-ray emitting electrons through diffusive shock acceleration and the flux decrease to synchrotron cooling of electrons, the observed time variabilities indicate that the magnetic field is amplified to $\sim 500~\mu{\rm G}$. 
The spectra of the stripes were found to be harder ($\Gamma=2.1\textrm{--}2.6$) than those of the rim ($\Gamma=2.7\textrm{--}2.9$), which would 
also be explained by amplified magnetic fields and/or stochastic acceleration in the stripes. 
Another finding is a tight anti-correlation between the surface brightness and photon indices of the stripe emission, which would indicate that 
only a small number of parameters control the temporal and spatial variation of the stripe emission. 
%--------------------------------------------------------------------------------

%--------------------------------------------------------------------------------
\begin{ack}

This work is  supported by JSPS/MEXT Scientific Research Grant Numbers 
 JP25109004 (T.T. and T.G.T.), JP19H01936  (T.T.),  JP26800102  (H.U.),  JP19K03915(H.U.), and JP15H02090 (T.G.T.).
\end{ack}

%%%
% See the manual for the detail.
%%%

\end{document}